\documentclass[aps,twocolumn,superscriptaddress,floatfix,amsmath,longbibliography,nofootinbib,amssymb]{revtex4-2}

\usepackage{graphicx}
\usepackage{dcolumn}
\usepackage{bm}
\usepackage{mathrsfs}
\usepackage{booktabs}
\usepackage{float}
\usepackage{xcolor}
\usepackage[colorlinks=true,citecolor=blue,linkcolor=blue,urlcolor=blue]{hyperref}

\newcommand{\ee}{\mathrm{e}}

\newcommand{\dd}{\mathrm{d}}
\newcommand{\Oc}{\langle\mathcal O_+\rangle}

\begin{document}

\title{Holographic subregion complexity in insulator/superconductor transition}

\author{Yu Shi}
\email{shiyu@hhtc.edu.cn}
\author{Chikun Ding}
\author{Yuebing Zhou}
\affiliation{Department of Physics, Huaihua University, Huaihua, Hunan 418008, China}

\author{Weike Deng}
\affiliation{School of Science, Hunan Institute of Technology, Hengyang 421002, China}

\author{Sheng Long}
\email{shenglong@ucas.ac.cn}
\thanks{Corresponding author}
\affiliation{School of Fundamental Physics and Mathematical Sciences, Hangzhou Institute for Advanced Study, University of Chinese Academy of Sciences, Hangzhou 310024, China}

\begin{abstract}
We study holographic subregion complexity (HSC) across a fully backreacted insulator/superconductor transition in an AdS-soliton background and compare it with holographic entanglement entropy (HEE) and holographic complexity based on the complexity=volume (CV) proposal. Both HSC and HEE signal the second-order transition. For a strip subsystem, competing connected and disconnected Ryu--Takayanagi surfaces give rise to a confinement/deconfinement transition. At fixed chemical potential in the superconducting phase, HSC exhibits a finite jump at the critical width, whereas HEE remains continuous. Beyond this width, HSC grows linearly with the strip width, while HEE is constant. At fixed strip width, HSC first decreases and then increases with chemical potential for $\ell<\ell_c$, opposite to HEE, but increases monotonically for $\ell>\ell_c$. After consistent normalization and subtraction of the respective insulating references, the half-space HSC and CV complexity densities are analytically identical. These results show that HSC can diagnose the insulator/superconductor transition, but its qualitative response remains sensitive to the subsystem scale and entanglement-wedge topology.
\end{abstract}

\maketitle

\section{Introduction}
\label{sec:introduction}

The AdS/CFT correspondence maps strongly coupled boundary dynamics to higher-dimensional gravity~\cite{Maldacena1998,GubserKlebanovPolyakov1998,Witten1998}. Holographic superconductors realize spontaneous $U(1)$ symmetry breaking through a charged condensate~\cite{Gubser2008,HartnollHerzogHorowitz2008PRL,HartnollHerzogHorowitz2008JHEP} and reproduce characteristic superconducting transport~\cite{Hartnoll2009,Herzog2009,Horowitz2011,CaiEtAl2015}. Backreaction, generalized interactions, and multicomponent condensates generate phase structures~\cite{KimKimSeo2025,ZhangEtAl2025} with competing or coexisting superconducting, charge, spin, nematic, and pair-density-wave orders in homogeneous and modulated settings~\cite{FradkinKivelsonTranquada2015,AgterbergEtAl2020,FlaugerEtAl2011,AmadoEtAl2014,KiritsisLi2016,CremoniniLiRen2017,LingWu2021}. A hairy solution alone does not identify the thermodynamically preferred branch, fix the transition order, or characterize the reorganization of the quantum state. Order parameters, transport coefficients, and the grand potential probe distinct aspects, while scalar hair does not establish thermodynamic dominance~\cite{FrancoEtAl2010}. Reliable probes of the transition and state reorganization are therefore central to holographic superconductivity.

Quantum-information observables provide natural nonlocal probes. Holographic entanglement entropy (HEE) maps boundary entanglement to extremal-surface area through the Ryu--Takayanagi (RT) prescription~\cite{RyuTakayanagi2006PRL,RyuTakayanagi2006JHEP,HubenyRangamaniTakayanagi2007,LewkowyczMaldacena2013}. In holographic superconductor models, HEE probes superconducting transitions and often distinguishes their order~\cite{AlbashJohnson2012,CaiEtAl2012,CaiPwave2012,KuangEtAl2014,YaoJing2014,PengPan2014,DeyMahapatraSarkar2014,PengLiu2017,DasFujitaKim2017,YaoEtAl2021,WangZhangYao2023,YaoCaiTian2026}, making it a benchmark for assessing other information measures.

Complexity characterizes quantum states differently from entanglement: entanglement entropy quantifies correlations, whereas circuit complexity counts the minimum elementary gates required to prepare a target state from a reference. The principal holographic proposals are complexity=volume (CV), based on a maximal codimension-one volume, and complexity=action (CA), based on the action of the Wheeler--DeWitt region~\cite{StanfordSusskind2014,Susskind2016,BrownEtAl2016PRL,BrownEtAl2016PRD}; their distinct geometric sensitivities have been studied in holographic superconductors~\cite{YangEtAl2019,AnLiYangYang2022}. For boundary subregions, subregion CV defines holographic subregion complexity (HSC) as the volume bounded by the subregion and its RT surface~\cite{Alishahiha2015,BenAmiCarmi2016}. Its field-theory dual remains unsettled: circuit and purification constructions do not uniquely reproduce the entanglement-wedge volume, while density-matrix Krylov complexity provides an alternative framework for mixed states~\cite{YangEtAl2022Dual,AgonHeadrickSwingle2019,AlishahihaBanerjee2023}. The ultraviolet structure of HSC and its responses to quenches, deformations, and confinement have also been studied~\cite{CarmiMyersRath2017,ReynoldsRoss2017,YangNiuKim2017,RoySarkar2017,ChenEtAl2018,LingLiuZhang2019,Zhang2019,JangEtAl2020}.

Previous studies show that HSC responds to metal/superconductor transitions in black-hole backgrounds, locating critical points and distinguishing transition orders~\cite{ShiEtAl2026,MomeniEtAl2016,ZangenehOngWang2017,Fujita2019,GuoKuangWang2019,Chakraborty2020,ShiPanJing2020,ShiPanJing2021,XuEtAl2023,WangEtAl2023,HuEtAl2024,HuEtAl2025,LaiPan2026}. Its behavior is less universal than that of HEE: the ordering of the normal and superconducting phases, temperature dependence, and even critical signatures can change with model parameters and subsystem scale~\cite{WangEtAl2023,HuEtAl2024,HuEtAl2025,LaiPan2026}. Which features of HSC then genuinely reflect changes in the superconducting system itself?

The AdS soliton provides a clean setting for this question. Its horizonless infrared cap describes a gapped insulator~\cite{NishiokaRyuTakayanagi2010,HorowitzWay2010}, while a strip admits connected and disconnected RT surfaces~\cite{CaiEtAl2012,YaoJing2014} whose area competition defines a confinement/deconfinement transition~\cite{KlebanovKutasovMurugan2008} alongside the thermodynamic insulator/superconductor transition. This geometry also permits direct comparison of finite-strip HSC, half-space HSC, and the CV complexity of the boundary pure state, isolating thermodynamic criticality, subsystem scale, and entanglement-wedge topology within one model.

In this work, we compute HSC in a fully backreacted AdS-soliton background and compare it with HEE and CV complexity. Both HSC and HEE signal the second-order insulator/superconductor transition, but HSC carries an additional scale dependence: its nonmonotonic response appears on the connected branch below the critical width, whereas the disconnected branch increases monotonically with chemical potential. At the confinement/deconfinement transition, HSC undergoes a finite jump associated with the change of RT branch, while HEE remains continuous. In the half-space limit, the finite HSC density is analytically identical to the finite CV density after common normalization and insulating-reference subtraction. These results separate the thermodynamic signal of HSC from features controlled by subsystem scale and entanglement-wedge topology.

The structure of this paper is organized as follows. Section~\ref{sec:transition} presents the backreacted model and transition, Sec.~\ref{sec:observables} studies strip and half-space HEE and HSC, and Sec.~\ref{sec:conclusion} concludes.

\section{Insulator/superconductor transition}
\label{sec:transition}

We consider the five-dimensional Einstein--Maxwell theory coupled to a charged complex scalar~\cite{CaiEtAl2012},
\begin{equation}
\begin{split}
 S={}&\int \dd^5x\sqrt{-g}\left[
 \mathcal R+\frac{12}{L^2}-\frac14F_{ab}F^{ab}\right.\\
 &\left.{}-|D\psi|^2-m^2|\psi|^2\right],
\end{split}
 \label{eq:action}
\end{equation}
where $L$ is the AdS radius, $F=\dd A$, and $D_a=\nabla_a-iqA_a$. Setting $L=1$, we incorporate the matter backreaction with the ansatz
\begin{equation}
\begin{aligned}
 \dd s^2={}&\frac{\dd r^2}{r^2B(r)}+r^2\left[
 -\ee^{C(r)}\dd t^2+\dd x^2+\dd y^2\right.\\
 &\left.{}+\ee^{A(r)}B(r)\dd\chi^2\right],\\[-2pt]
 A_a\dd x^a={}&\phi(r)\dd t,\qquad \psi=\psi(r).
\end{aligned}
 \label{eq:metric-r}
\end{equation}
The radial coordinate ranges from the soliton tip $r=r_0$ to the AdS boundary. Requiring $B(r_0)=0$ and removing the conical singularity fixes the periodicity of $\chi$:
\begin{equation}
 \chi\sim\chi+\Gamma,\qquad
 \Gamma=\frac{4\pi\ee^{-A(r_0)/2}}{r_0^2B'(r_0)}.
 \label{eq:period}
\end{equation}

\begin{figure*}[t!]
\centering
\begin{minipage}[t]{0.475\textwidth}
\centering
\includegraphics[width=\linewidth]{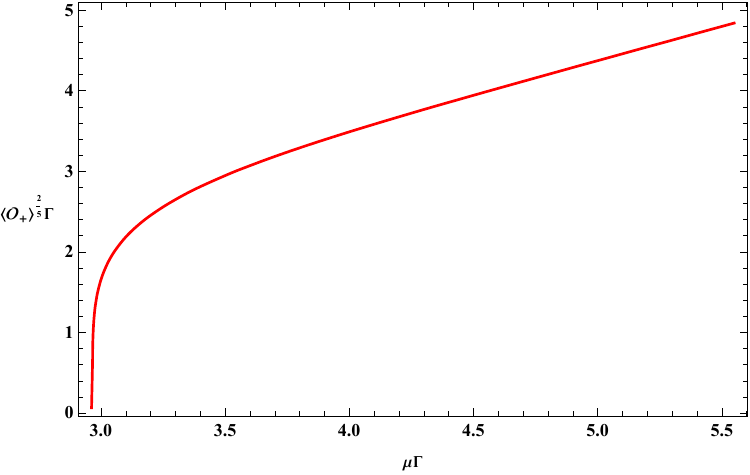}
\end{minipage}\hfill
\begin{minipage}[t]{0.475\textwidth}
\centering
\includegraphics[width=\linewidth]{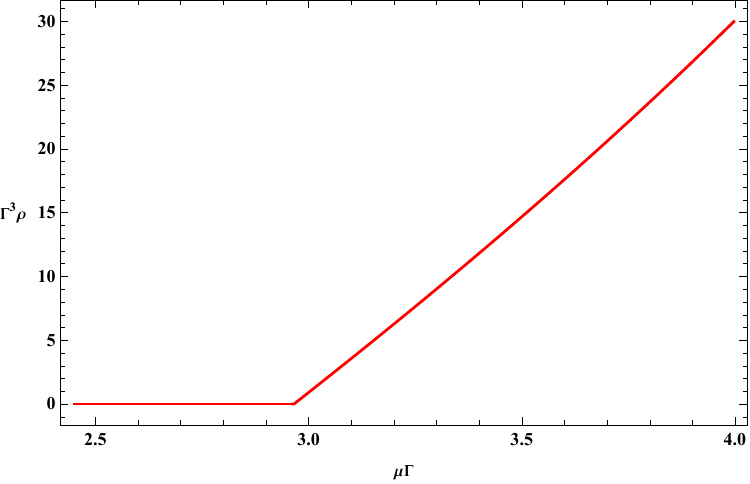}
\end{minipage}
\caption{Source-free condensate $\Oc^{2/5}\Gamma$ (left) and charge density $\rho\Gamma^3$ (right) versus $\mu\Gamma$.}
\label{fig:transition}
\end{figure*}

With a prime denoting $\dd/\dd r$, the scalar and Maxwell equations are
\begin{align}
 0={}&\psi''+\left(\frac5r+\frac{A'}2+\frac{B'}B+\frac{C'}2\right)\psi'
 \nonumber\\
 &+\frac1{r^2B}\left(\frac{q^2\phi^2}{\ee^C r^2}-m^2\right)\psi,
 \label{eq:eom-psi}\\
 0={}&\phi''+\left(\frac3r+\frac{A'}2+\frac{B'}B-\frac{C'}2\right)\phi'
 \nonumber\\
 &-\frac{2q^2\psi^2}{r^2B}\phi.
 \label{eq:eom-phi}
\end{align}
The three independent Einstein equations read
\begin{align}
 A'={}&\frac{1}{r(6+rC')}\Bigl(2r^2C''+r^2C'^2+4rC'
 \nonumber\\[-2pt]
 &\hspace{4.4em}{}+4r^2\psi'^2-2\ee^{-C}\phi'^2\Bigr),
 \label{eq:eom-A}\\
 0={}&C''+\frac{C'^2}{2}+\left(\frac5r+\frac{A'}2+\frac{B'}B\right)C'
 \nonumber\\[-2pt]
 &{}-\frac{\ee^{-C}}{r^2}\left(\phi'^2+\frac{2q^2\phi^2\psi^2}{r^2B}\right),
 \label{eq:eom-C}\\
 0={}&B'\left(\frac3r-\frac{C'}2\right)
 +B\biggl(\psi'^2-\frac{A'C'}2+\frac{\ee^{-C}\phi'^2}{2r^2}
 \nonumber\\[-2pt]
 &\hspace{4.4em}{}+\frac{12}{r^2}\biggr)
 +\frac1{r^2}\biggl(\frac{\ee^{-C}q^2\phi^2\psi^2}{r^2}
 \nonumber\\[-2pt]
 &\hspace{4.4em}{}+m^2\psi^2-12\biggr),
 \label{eq:eom-B}
\end{align}

A regular expansion at $r=r_0$, with $B(r_0)=0$, leaves four independent tip data: $r_0$, $\psi(r_0)$, $\phi(r_0)$, and $C(r_0)$. The equations are invariant under the following scalings, where $\boldsymbol{\xi}=(\chi,x,y,t)$:
\begin{align}
 (r,\boldsymbol{\xi},\phi)&\to
 (\alpha r,\alpha^{-1}\boldsymbol{\xi},\alpha\phi),
 \label{eq:scaling-r}\\[-2pt]
 (C,t,\phi)&\to
 (C-2\ln\beta,\beta t,\phi/\beta).
 \label{eq:scaling-t}
\end{align}
We therefore construct seed solutions with $r_0=1$ and $C(r_0)=0$. After integration, the boundary coordinates are rescaled to impose $A(\infty)=C(\infty)=0$, while asymptotically AdS behavior requires $B(\infty)=1$. The matter fields behave near the AdS boundary as
\begin{align}
 \phi(r)&=\mu-\frac{\rho}{r^2}+\cdots,
 \label{eq:asymptotic-phi}\\[-2pt]
 \psi(r)&=\frac{\psi_-}{r^{\Delta_-}}
 +\frac{\psi_+}{r^{\Delta_+}}+\cdots .
 \label{eq:asymptotic-psi}
\end{align}
Here $\Delta_\pm=2\pm\sqrt{4+m^2}$ are the conformal dimensions, while $\mu$ and $\rho$ are the chemical potential and charge density. We take $m^2=-15/4$ and $q=2$, impose $\psi_-=0$, and identify $\Oc=\psi_+$.

For fixed $\psi(r_0)$, we solve the coupled equations in $z=r_0/r$ by the standard shooting method, tuning $\phi(r_0)$ to impose the source-free condition $\psi_-=0$. Results are reported in terms of $\mu\Gamma$, $\rho\Gamma^3$, $\Oc^{2/5}\Gamma$, and $\ell/\Gamma$. The source-free hairy branch bifurcates from the insulating solution at $\mu_c\Gamma=2.9662$. Figure~\ref{fig:transition} shows that the condensate and charge density turn on continuously at this point, a characteristic signature of a second-order insulator/superconductor transition.

\section{Entanglement entropy and subregion complexity}
\label{sec:observables}

The HEE of a boundary subregion $\mathcal A$ on a static time slice is given by the RT prescription~\cite{RyuTakayanagi2006PRL},
\begin{equation}
 S_{\mathcal A}=\frac{\operatorname{Area}(\gamma_{\mathcal A})}{4G_N}.
 \label{eq:rt-prescription}
\end{equation}
For the corresponding HSC, we adopt the subregion-CV normalization~\cite{Alishahiha2015},
\begin{equation}
 \mathcal C_{\mathcal A}=\frac{\operatorname{Vol}(\Sigma_{\mathcal A})}{8\pi L G_N}.
 \label{eq:hsc-prescription}
\end{equation}
Here $G_N$ is Newton's gravitational constant and $L$ is the AdS radius; for the five-dimensional model below, $G_N=G_5$. The codimension-two minimal surface $\gamma_{\mathcal A}$ is anchored on $\partial\mathcal A$ and homologous to $\mathcal A$, while $\Sigma_{\mathcal A}$ is the codimension-one region bounded by $\mathcal A$ and $\gamma_{\mathcal A}$. The minimum-area RT surface is physical, and HSC is evaluated for the wedge selected by this surface.

\subsection{Strip subsystem}

For a strip $-\ell/2\leq x\leq\ell/2$, with regulated length $R$ along $y$ and period $\Gamma$ along $\chi$, we parametrize the RT surface $\gamma_{\mathcal A}$ on a constant-time slice by $x=x(z)$. Its induced metric is
\begin{equation}
 \begin{aligned}
 \dd s_{\gamma_{\mathcal A}}^2={}&\frac1{z^2}\left[
 \left(\frac1{B(z)}+x'(z)^2\right)\dd z^2+\dd y^2\right.\\[-2pt]
 &\left.\hspace{5.2em}{}+B(z)\ee^{A(z)}\dd\chi^2\right].
 \end{aligned}
 \label{eq:induced-metric}
\end{equation}
Let $z_*$ denote the maximal bulk depth. By reflection symmetry, it is sufficient to consider the $x\geq0$ half of the RT surface, for which $x(z_*)=0$ and $\lim_{\epsilon\to0}x(\epsilon)=\ell/2$. Writing $h(z)=\ee^{A(z)/2}$, the RT area and the enclosed volume give
\begin{align}
 S_{\mathcal A}&=\frac{R\Gamma}{2G_5}
 \int_\epsilon^{z_*}\frac{h(z)}{z^3}
 \sqrt{1+B(z)x'(z)^2}\,\dd z
 \nonumber\\[-2pt]
 &\hspace{1.8em}{}
 =\frac{R\Gamma}{4G_5}\left(s+\frac1{\epsilon^2}\right),
 \label{eq:hee-integral}\\
 \mathcal C_{\mathcal A}&=\frac{R\Gamma}{4\pi L G_5}
 \int_\epsilon^{z_*}\frac{h(z)x(z)}{z^4}\,\dd z
 \nonumber\\[-2pt]
 &\hspace{1.8em}{}
 =\frac{R\Gamma}{8\pi L G_5}
 \left(c+\frac{\ell}{3\epsilon^3}\right).
 \label{eq:hsc-integral}
\end{align}
Here $\epsilon$ is the ultraviolet cutoff, while $s$ and $c$ denote the finite parts of HEE and HSC, respectively, after the displayed ultraviolet divergences have been subtracted. Defining $X(z)=B(z)\ee^{A(z)}/z^6$, translational invariance along $x$ yields
\begin{equation}
 p(z;z_*)\equiv |x'(z)|
 =\left[B(z)\left(\frac{X(z)}{X(z_*)}-1\right)\right]^{-1/2}.
 \label{eq:strip-slope}
\end{equation}
The strip width and the surface profile are then
\begin{equation}
 \ell(z_*)=2\int_0^{z_*}p(z;z_*)\dd z,
 \qquad x(z;z_*)=\int_z^{z_*}p(u;z_*)\dd u.
 \label{eq:strip-profile}
\end{equation}

\begin{figure*}[!t]
\centering
\begin{minipage}[t]{0.475\textwidth}
\centering
\includegraphics[width=\linewidth]{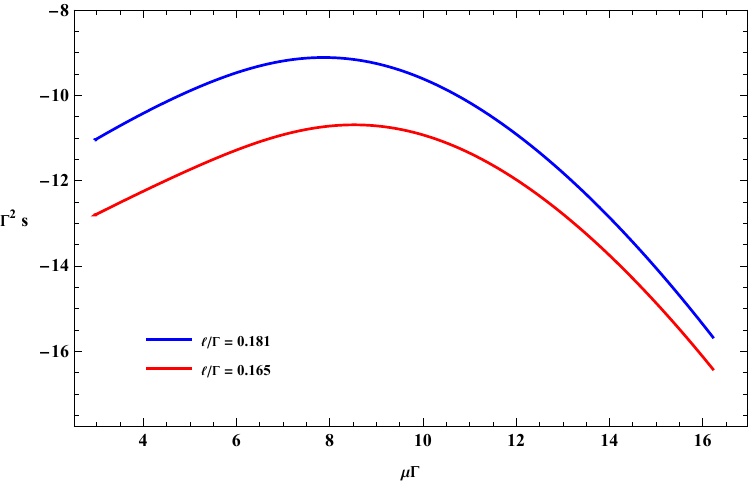}
\end{minipage}\hfill
\begin{minipage}[t]{0.475\textwidth}
\centering
\includegraphics[width=\linewidth]{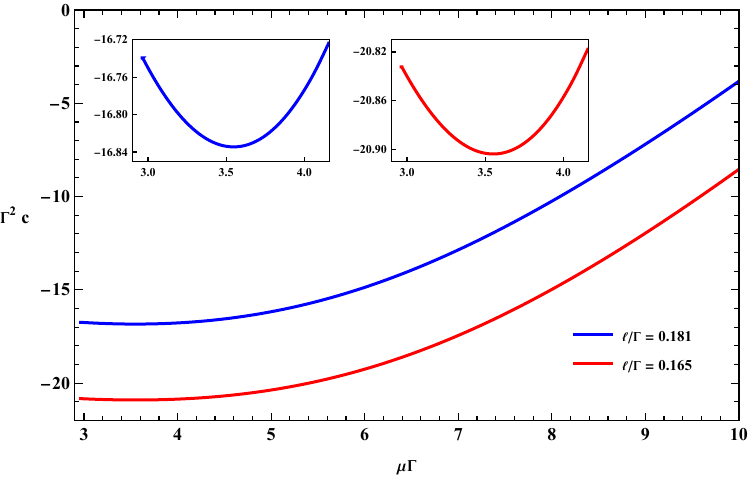}
\end{minipage}
\caption{Finite HEE (left) and HSC (right) versus $\mu\Gamma$ at $\ell/\Gamma=0.165$ and $0.181$.}
\label{fig:fixed-strips}
\end{figure*}

\begin{samepage}
Exchanging the two radial integrations implicit in $x(z;z_*)$ gives the following equivalent form for the finite part of Eq.~\eqref{eq:hsc-integral}:
\begin{equation}
 c=2\int_0^{z_*}
 \frac{[h(z)-1]x(z;z_*)-\frac13z\,p(z;z_*)}{z^4}\dd z,
 \label{eq:finite-c}
\end{equation}
The cubic ultraviolet divergence now cancels at the integrand level, making this representation numerically more stable.
\end{samepage}

\begin{figure*}[!t]
\centering
\begin{minipage}[t]{0.475\textwidth}
\centering
\includegraphics[width=\linewidth]{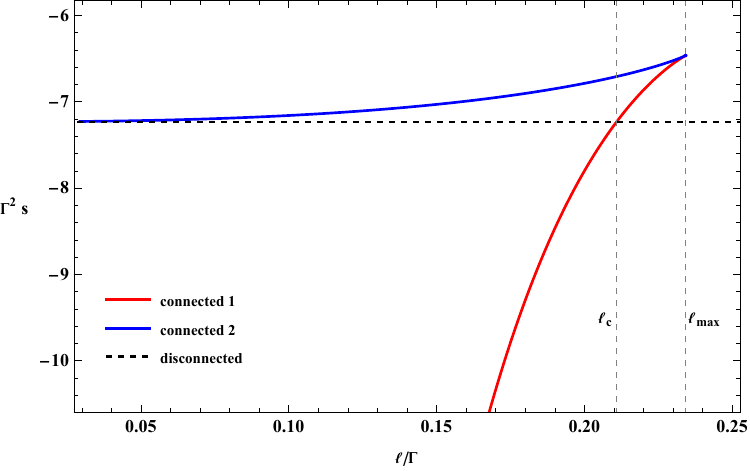}
\end{minipage}\hfill
\begin{minipage}[t]{0.475\textwidth}
\centering
\includegraphics[width=\linewidth]{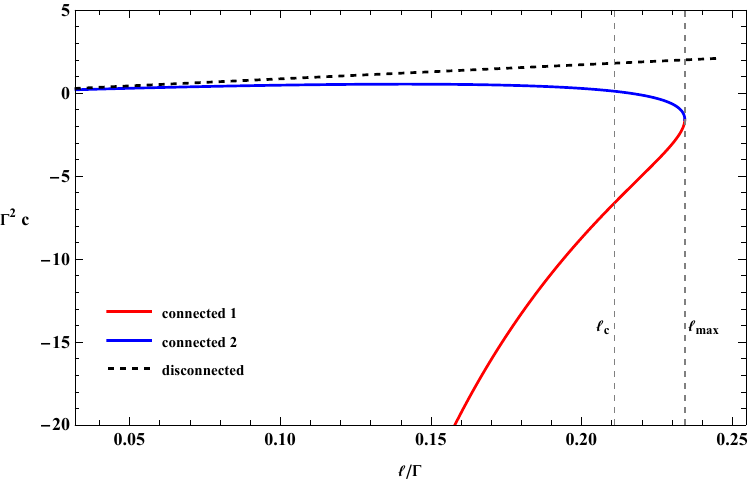}
\end{minipage}
\caption{Finite HEE (left) and HSC (right) versus $\ell/\Gamma$ at $\mu\Gamma=6.9165$. The vertical dashed lines mark $\ell_c$ and $\ell_{\max}$.}
\label{fig:width-branches}
\end{figure*}

We first examine the thermodynamic transition at fixed subsystem size. Figure~\ref{fig:fixed-strips} shows HEE and HSC at $\ell/\Gamma=0.165$ and $0.181$. For $\mu<\mu_c$, the $\mu$-independent insulating geometry keeps both quantities constant. At both widths, the physical RT surface remains smooth, and both observables depart continuously from their insulating values at $\mu_c$, as expected for a second-order transition. In the superconducting phase, HEE first rises and then falls, a recurring feature of holographic insulator/superconductor models~\cite{CaiEtAl2012,YaoJing2014,PengPan2014}. This behavior has been attributed to a reorganization of low-energy degrees of freedom~\cite{CaiEtAl2012}: initial condensation may release mobile-carrier or quasiparticle modes and raise HEE, whereas stronger condensation incorporates more degrees of freedom into the ordered state and lowers it. HSC shows the opposite trend, reaching a minimum at a different chemical potential from the HEE maximum.

We next vary the subsystem scale at $\mu\Gamma=6.9165$. Figure~\ref{fig:width-branches} shows three candidates below $\ell_{\max}$: the connected-1, connected-2, and disconnected branches; connected solutions cease to exist above $\ell_{\max}$. The minimum-area criterion selects the physical RT surface. Connected-1 dominates for narrow strips, whereas the $\ell$-independent disconnected HEE intersects it at $\ell_c/\Gamma=0.2109$ and dominates thereafter. Thus, $\ell_c$ defines the confinement/deconfinement transition, while $\ell_{\max}$ marks only the endpoint of connected solutions. The disconnected HSC grows linearly with $\ell$. Although the competing saddles have equal areas at $\ell_c$, their finite volumes differ, so the physical HSC jumps when the RT branch changes.

Similar volume discontinuities in confining geometries and holographic QCD models arise from changes in the topology of the physical RT surface~\cite{BenAmiCarmi2016,Zhang2019,Nakajima2026}. The same mechanism persists here in a fully backreacted superconducting soliton background. The HSC jump at $\ell_c$ therefore characterizes entanglement-wedge reconstruction at the confinement/deconfinement transition, rather than a new insulator/superconductor transition. Together, the thermodynamic boundary $\mu=\mu_c$ and scale-dependent boundary $\ell=\ell_c(\mu)$ divide the $(\mu,\ell)$ plane into four regions: confined insulator, deconfined insulator, confined superconductor, and deconfined superconductor~\cite{CaiEtAl2012,YaoJing2014}. The first boundary separates the insulating and superconducting phases, whereas the second distinguishes the two RT-surface topologies, defining the combined thermodynamic and RT-topological phase structure.

\begin{figure*}[t]
\centering
\begin{minipage}[t]{0.475\textwidth}
\centering
\includegraphics[width=\linewidth]{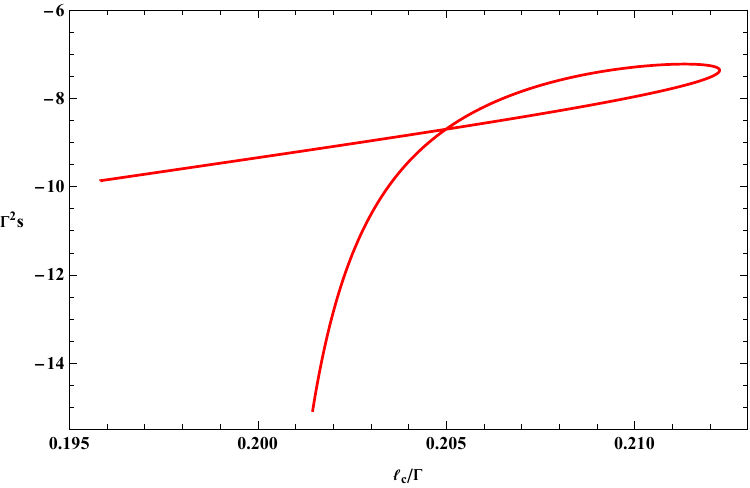}
\end{minipage}\hfill
\begin{minipage}[t]{0.475\textwidth}
\centering
\includegraphics[width=\linewidth]{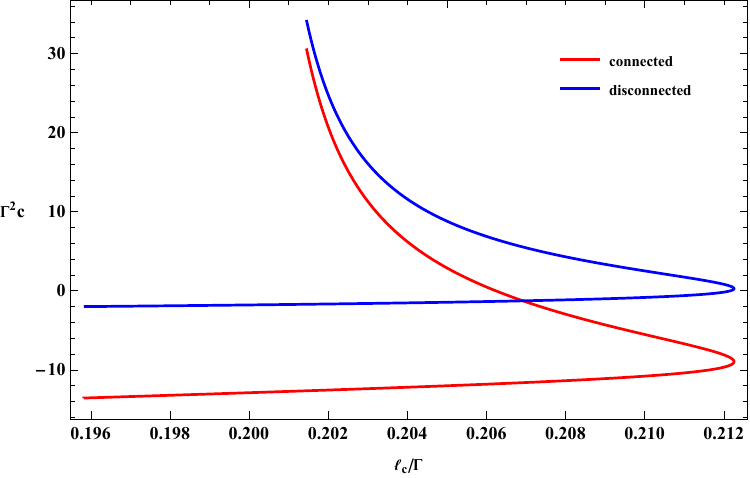}
\end{minipage}
\caption{Finite HEE (left) and HSC (right) along the critical-width curve $\ell_c(\mu)$.}
\label{fig:critical-values}
\end{figure*}

Figure~\ref{fig:critical-values} extends the comparison over the full critical-width curve. At each $\mu$, the competing saddles have equal $s$ at $\ell_c(\mu)$, so the left panel shows one HEE value. As $\mu$ varies, these values form a self-intersecting curve. Their finite, unequal $c$ values give two HSC branches in the right panel, although the projected curves may cross. The area crossing locates the confinement/deconfinement boundary, while the HSC gap quantifies the corresponding volume reconstruction.

The strip results show that HSC responds both to the second-order insulator/superconductor transition and to RT-surface reconstruction at the confinement/deconfinement transition. Its nonmonotonic behavior at small widths, however, is not scale independent: studies of metal/superconductor transitions also find that the qualitative HSC response varies with strip width~\cite{ShiEtAl2026,Fujita2019,XuEtAl2023}, and the present results provide evidence for the same dependence in this insulator/superconductor system. We therefore turn to the half-space subsystem as a natural large-scale probe and compare its HSC with CV complexity.

\subsection{Half-space subsystem}

The half-space is bounded by a planar RT sheet extending from the boundary to the soliton tip. Let $L_x$ be the regulated length in the noncompact $x$ direction. Its HEE and HSC are
\begin{align}
 S_H&=\frac{R\Gamma}{4G_5}\int_\epsilon^1\frac{h(z)}{z^3}\dd z
 =\frac{R\Gamma}{8G_5}\left(s+\frac1{\epsilon^2}\right),
 \nonumber\\[2pt]
 \mathcal C_H&=\frac{R\Gamma L_x}{8\pi L G_5}\int_\epsilon^1\frac{h(z)}{z^4}\dd z
 =\frac{R\Gamma L_x}{8\pi L G_5}
 \left(\bar c+\frac1{3\epsilon^3}\right).
 \label{eq:half-sc}
\end{align}
Here $\bar c$ is the finite HSC density per unit length in the $x$ direction. We display the latter as the dimensionless quantity $\Gamma^3\bar c$. In the insulating phase, $h(z)=1$, giving $\Gamma^2s=-\pi^2$ and $\Gamma^3\bar c=-\pi^3/3$.

\begin{figure*}[t]
\centering
\begin{minipage}[t]{0.475\textwidth}
\centering
\includegraphics[width=\linewidth]{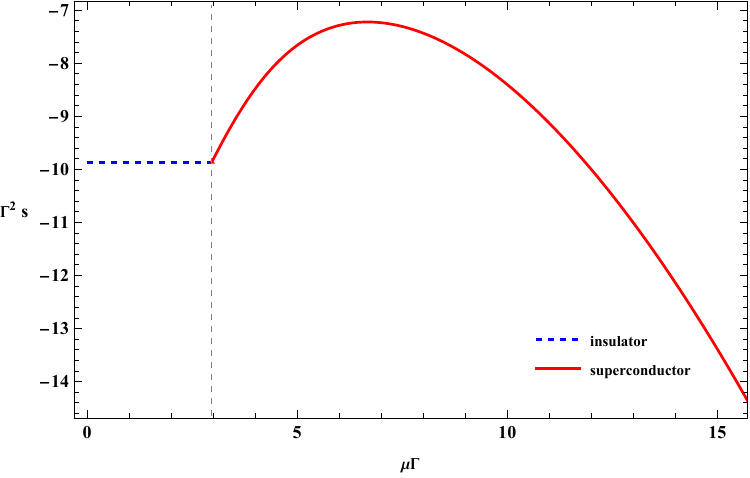}
\end{minipage}\hfill
\begin{minipage}[t]{0.475\textwidth}
\centering
\includegraphics[width=\linewidth]{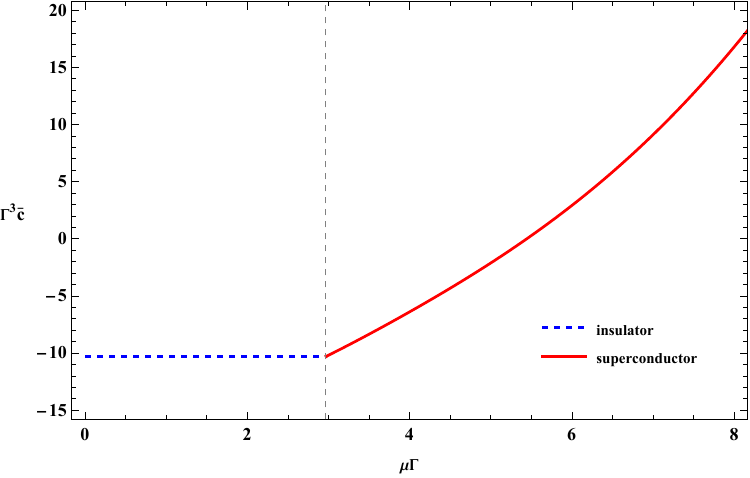}
\end{minipage}
\caption{Finite half-space HEE (left) and HSC density (right) versus $\mu\Gamma$. The vertical dashed line marks $\mu_c\Gamma=2.9662$.}
\label{fig:half}
\end{figure*}

Figure~\ref{fig:half} shows that both HEE and HSC retain a clear response at $\mu_c$ and hence signal the insulator/superconductor transition. In the superconducting phase, HEE remains nonmonotonic, whereas HSC increases monotonically. The minimum found for the narrow strips in Fig.~\ref{fig:fixed-strips} is absent in the half-space geometry, directly showing the scale dependence of HSC.

Previous work suggests that, as the subsystem size increases, the qualitative response of HSC can approach that of pure-state CV complexity~\cite{ShiEtAl2026}. HSC is commonly regarded as a candidate probe of reduced mixed-state complexity, although no unique field-theory interpretation has been established~\cite{AgonHeadrickSwingle2019}. By contrast, CV complexity here refers to the boundary ground state, so their boundary interpretations differ. The present model nevertheless permits a direct geometric comparison. For CV complexity, we adopt the conventional normalization~\cite{StanfordSusskind2014},
\begin{equation}
 \mathcal C_V=\frac{\operatorname{Vol}(\mathcal B)}{G_N L},
 \label{eq:cv-definition}
\end{equation}
where $\mathcal B$ is the maximal codimension-one bulk surface anchored on the complete boundary time slice. Denoting the regulated area of the noncompact $(x,y)$ directions by $V_{xy}$, the CV density in the present static geometry is
\begin{equation}
 \frac{\mathcal C_V}{V_{xy}}
 =\frac{\Gamma}{G_5L}\int_\epsilon^1\frac{h(z)}{z^4}\dd z.
 \label{eq:cv-density}
\end{equation}
This expression shows that the CV and half-space HSC densities have the same radial kernel. We define the corresponding dimensionless densities and their reference-subtracted versions by
\begin{equation}
\begin{aligned}
 \widehat{\mathcal C}_H&\equiv
 \frac{8\pi L G_5\Gamma^2}{R L_x}\mathcal C_H,
 &\widehat{\mathcal C}_V&\equiv
 \frac{G_5L\Gamma^2}{V_{xy}}\mathcal C_V,\\[-2pt]
 \Delta\widehat{\mathcal C}_i&\equiv
 \widehat{\mathcal C}_i-\widehat{\mathcal C}_{i,\mathrm{ins}},
 &i&=H,V.
\end{aligned}
\label{eq:normalized-complexities}
\end{equation}
Here ``ins'' denotes the pure AdS-soliton insulating reference with the same physical period $\Gamma$. The two reference-subtracted densities therefore satisfy
\begin{equation}
 \Delta\widehat{\mathcal C}_V=\Delta\widehat{\mathcal C}_H
 =\Gamma^3\bar c+\frac{\pi^3}{3}.
 \label{eq:cv-half}
\end{equation}

In the horizonless AdS soliton, the maximal CV slice reaches the smooth tip~\cite{ReynoldsRoss2018Soliton}. The half-space wedge reaches the same tip, giving Eq.~\eqref{eq:cv-half} its direct geometric origin. Neither the half-space geometry nor the large-subsystem limit, however, generally implies an equivalence between HSC and CV complexity. In geometries with horizons, the HSC integration region remains bounded by the corresponding RT/HRT surface, whereas the maximal CV slice can extend through the horizon and probe the black-hole interior~\cite{AnLiYangYang2022,AgonHeadrickSwingle2019}. Equation~\eqref{eq:cv-half} is therefore specific to the present static, horizonless soliton geometry and the use of a common reference subtraction; it should not be expected to hold for generic finite subregions or black-hole backgrounds.

\section{Discussion and conclusion}
\label{sec:conclusion}

We investigated HSC across a fully backreacted insulator/superconductor transition in an AdS-soliton background and compared it with HEE and CV complexity. Both HEE and HSC clearly diagnose the second-order transition. At the strip confinement/deconfinement transition, HEE remains continuous, whereas HSC exhibits a finite jump induced by the change of the physical RT surface. The response of HSC to the chemical potential is scale dependent: it first decreases and then increases for $\ell<\ell_c$, but increases monotonically for $\ell>\ell_c$. We further found that, after the appropriate normalization and subtraction of the insulating reference, the half-space HSC density and the CV complexity density are analytically identical.

For the specific purpose of probing holographic superconducting phase transitions, we find HEE better suited than HSC. First, both observables respond equally clearly to the second-order critical point in the present model, and HSC does not provide a more robust diagnostic than HEE. Second, the critical width $\ell_c$ is fixed by the equal-area condition for the competing RT surfaces and is therefore determined directly by HEE. The finite jump in HSC is a consequence of the associated entanglement-wedge volume reconstruction and does not independently determine $\ell_c$. Finally, HSC exhibits pronounced scale dependence. Small subregions predominantly probe the near-boundary geometry, where both HEE and HSC generally retain clear critical responses. As the subsystem size increases, HSC assigns greater weight to the deep-infrared geometry: the near-horizon exterior in black-hole backgrounds and the vicinity of the soliton tip in the present horizonless geometry. Here the nonmonotonic behavior of HSC occurs only on the connected branch with $\ell<\ell_c$ and disappears for $\ell>\ell_c$. This scale dependence can be even more pronounced in metal/superconductor transitions, where the ability of HSC to diagnose the transition may be weakened at some widths~\cite{ShiEtAl2026,Fujita2019,XuEtAl2023}. The diverse behavior of HSC may therefore reflect competition between the ultraviolet and infrared geometries and between different RT branches, rather than necessarily signaling new changes in the superconducting phase structure.

These observations do not diminish the value of the HSC prescription. Rather, its scale dependence and nontrivial volume response reveal a fundamental difference between the information encoded by complexity and by entanglement entropy. One of the central motivations for holographic complexity is to probe black-hole interiors. In a top-down M-theory holographic superconductor, the CA complexity growth rate is discontinuous at the critical temperature and is sensitive to the accompanying abrupt reconstruction of the interior geometry~\cite{AnLiYangYang2022}. The recently proposed timelike holographic subregion complexity (THSC) extends subregion CV to Lorentzian timelike regions and allows the relevant extremal surfaces to cross the horizon, providing a new geometric route for subregion complexity to probe black-hole interiors~\cite{AlishahihaTimelike2025}. Applying THSC to holographic superconducting black holes and testing whether it can identify condensate-induced reconstruction of the interior geometry is therefore a natural direction that more closely follows the original motivation for holographic complexity.

\begin{acknowledgments}
This work was supported by the National Natural Science Foundation of China under Grant Nos.~12305063, 12547143, and 12375047, by the Hunan Provincial Natural Science Foundation of China under Grant No.~2026JJ30135, and by the Scientific Research Fund of Hunan Provincial Education Department under Grant No.~25B0823.
\end{acknowledgments}

\begingroup
\renewcommand{\bibfont}{\scriptsize\setlength{\baselineskip}{8.2pt}}
\setlength{\emergencystretch}{1em}

\endgroup

\end{document}